\documentclass[12pt,preprint]{aastex}

\begin{document}
\def\ch{{\bf changed}\ }

\title{Using All-Sky Surveys to Find Planetary Transits}

\author{Joshua Pepper, Andrew Gould and  D.\ L.\ DePoy}
\affil{Department of Astronomy, Ohio State University, Columbus, Ohio 43210}
\email{pepper@astronomy.ohio-state.edu, gould@astronomy.ohio-state.edu, 
depoy@astronomy.ohio-state.edu}

\begin{abstract}

Transits of bright stars offer a unique 
opportunity to study detailed properties of extrasolar planets that 
cannot be determined through radial-velocity observations.  We propose a new 
technique to find such systems using all-sky 
small-aperture transit surveys.  We 
derive a general formula for the number of stars that can be probed for 
such systems as a function of the characteristics of the star, the planet, 
and the survey.  We use this formula to derive the optimal telescope
design for finding transits of bright stars: a 5 cm ``telescope''
with a $4k\times 4k$ camera.

\end{abstract}


\section{INTRODUCTION}

In the past three years, great strides have been made in the detection of
extrasolar planets (XSPs).  To date, nearly all of the roughly 100 known 
XSPs have been discovered using the radial 
velocity (RV) technique.  However, RV detections, in and of themselves, yield 
only a few planetary parameters, namely the period $P$, the eccentricity $e$, 
and $M \sin (i)$, where $M$ is the mass of the planet and $i$ is the inclination 
of its orbit.  By contrast, if a planet transits its host star, much more 
information is available.  First, of course, the $M \sin (i)$ 
degeneracy can be broken.  Second, the ratio of the radii of the planet and 
host star can be measured.  Therefore, provided that the star can be 
classified well enough to determine its mass and radius, 
then the planet's radius and hence its density 
can be determined.  Third, and perhaps most important, if the transits can 
be  observed with sufficient signal-to-noise ratio (S/N), then one can probe 
otherwise unobservable details of a planet, such as its oblateness \citep{hui02}, 
atmospheric conditions \citep{char02}, and perhaps satellites and rings.  
Regardless of how a planet is initially discovered, once it is determined 
to transit its host star, this wealth of information can in principle be 
extracted by intensive follow-up observation of these transits.  This fact 
has been amply demonstrated by the discovery and analysis of the transiting 
planet HD209458b \citep{char00, cody02}.

At the moment, all ongoing and proposed transit surveys are carried out in 
relatively narrow pencil beams.  They make up for their small angular area 
with relatively deep exposures.  These surveys fall into two basic classes: 
field stars 
\citep{how00, brown99, mal01, udal02}\footnote{
\url{http://www.psi.edu/\~{}esquerdo/asp/asp.html}
\\
\url{http://www.hao.ucar.edu/public/research/stare/stare.html}
\\
\url{http://bulge.princeton.edu/\~{}ogle/}}, 
and clusters (\citealt{str00}; \citealt{burke02})\footnote{
\url{http://star-www.st-and.ac.uk/\~{}yt2/WEB\_GROUP/top.html}}.  These 
surveys are potentially 
capable of establishing the frequency of planets in various environments, 
but they are unlikely to find the kinds of transits of bright stars that would 
be most useful for intensive follow-up analysis.  
Although some of the surveys
of field stars are considered ``wide field'', their total survey areas are 
small compared to $4\pi$ str.  One project that has the 
potential to cover a very large area is WASP \citep{str02}, which plans to
employ 
five cameras, each with a $9.^{\circ}5 \times 9.^{\circ}5$ field of view.

An alternative method is to conduct an all-sky survey.  Instead of continuous
observation of all targets (which is impossible from a practical
standpoint for an all-sky survey), this approach would necessarily 
involve revisiting each target in the sky at regular, semi-regular, or random 
intervals throughout the course of the project.  
This kind of observing strategy will not yield a continuous light curve 
on any star, as the current transit surveys do.  Rather, this plan will 
generate an only sporadically sampled light curve.  However, the long time 
baseline for the survey will eventually generate just as many individual 
observations of a single star.  Transit-like dips in the data stream will 
not be visually obvious, but by repeatedly phase-folding the full light 
curve back on itself over a range of periods, one can detect the dips from 
the transits. (See \S~\ref{sec:randomnoise}.)

This approach is especially relevant given the fact that there are several
all-sky surveys already being planned for objectives other than transit
detections.  It should be possible, for instance, to utilize the
photometric data stream of upcoming astrometric missions for 
transit detection.  Space-based projects such as
GAIA\footnote{http://astro.estec.esa.nl/GAIA/gaia.html} and
DIVA\footnote{http://www.ari.uni-heidelberg.de/diva/diva.html} would
take hundreds of observations of millions of stars over mission lengths
of years with the aim of obtaining precise astrometry.  These data could
equally well be analyzed for planetary transits.

There are also several existing or proposed ground-based all-sky surveys, 
including the Large-aperture Synoptic Survey Telescope
(LSST)\footnote{http://www.lssto.org/lssto/index.htm}, the Panoramic
Optical Imager (POI), and the All-Sky Automated Survey
(ASAS)\footnote{http://www.astrouw.edu.pl/\~{}gp/asas/asas.html} \citep{poj00}.  
These surveys will be
imaging the entire sky every few days, a qualitatively similar cadence to
those of GAIA and DIVA.  While LSST and POI will likely be saturated  
by stars of $V\la 12$, which are of greatest interest for transit follow-ups,
ASAS is of particular interest in the present context because of its very
small aperture.

In this paper, we examine the process of analyzing photometric data
streams from all-sky surveys to find transits.  We calculate the sensitivity 
to XSP detection, the distance out to which these detections will be possible,
and the number of false-positive detections due to random noise.  We 
derive a general expression for the number of stars that can be probed 
by this technique as a function of the total sensitivity of the survey.  
We apply our analysis to the problem of telescope design and conclude
that very small, 5 cm telescopes are optimal for finding transits of bright
stars.

\section{SCALING RELATIONS} \label{sec:scalingrelations}

In order to accurately determine the number of transits a given survey would be
expected to detect, we must carefully define the set of stellar systems that 
can be probed for transits by the survey.  This will evidently depend on the 
number density $n$, the luminosity $L$, and the radius $R$, of the stars being 
probed; as well as the semi-major axis $a$ and radius $r$ of their planets.  A 
quick (and naive) formulation would then state that the total number of systems 
that can be probed for transits $N_{p}$ is  
\begin{equation} \label{equNt0}
N_{p} = \frac{4}{3} \pi n \frac{R}{a} \, [d_{max}(L,R,a,r)]^{3}  
\end{equation}
for a homogeneous population of such stars, where $d_{max}$ is the distance out to 
which a transit can be detected.

However, the quantity $d_{max}$ is ill-defined.  For fixed $L$, $R$, $a$, $r$ 
and distance $d$, the transit detection scales as $(1-x^{2})^{1/2}$ where $x$ is 
the transit impact parameter normalized to $R$; that is, $0 \leq x \leq 1$.  Hence, 
the signal-to-noise ratio scales as $(1-x^{2})^{1/4}$.  Since detections normally 
require a minimum S/N, $d_{max}$ must also be a function of the impact parameter 
$x$.  Thus, for a photon-noise limited survey, $d_{max} \propto (1-x^{2})^{1/4}$, 
and therefore
\begin{equation} \label{equNt1}
\frac{dN_{p}}{dx} = \frac{4}{3} \pi \, n \frac{R}{a} [d_{max}(L,R,a,r,x)]^{3}
\end{equation}
where $d_{max}(L,R,a,r,x) = d_{max}(L,R,a,r,0)(1-x^{2})^{1/4}$.  
Here $d_{max}(L,R,a,r,0)$ is 
the distance out to which a transit can be detected for an edge-on 
($i$ = $90^{\rm o}$) orbit.  We must then integrate over all values of the 
impact parameter $x$ from 0 to 1, and so
\begin{equation} \label{equNt}
N_{p} = \int \frac{dN_{p}}{dx} = \frac{4 \pi n R}{3 a} \, \eta [d_{max}(L,R,a,r,0)]^{3}
\end{equation}
where,
\begin{equation} \label{equeta}
\eta \, = \int_0^1 (1 - x^2)^{3/4} dx \: = \,
\frac{\sqrt{\pi}}{2}\frac{\Gamma(7/4)}{\Gamma(9/4)} \, \approx \, 0.719 \:.
\end{equation}
We now determine the dependence of $N_{p}$, the total number of systems 
probed, on the remaining parameters $L$, $R$, $a$, and $r$.  To do so, we 
analyze the detection requirement,
\begin{equation} \label{equchi}
N_{t} \left( \frac{\delta}{\sigma} \right)^{2} \ge \Delta \chi_{\rm min}^{2}\, ,
\end{equation}
where $N_{t}$ is the number of observations of the transit over the length 
of the survey, $\delta$ is
the fractional change in the star's brightness during the transit,
$\sigma$ is the fractional error of an individual flux measurement, and
$\Delta \chi_{\rm min}^{2}$ is the minimum
acceptable
difference in $\chi^{2}$ between a fit that assumes a constant flux and
one that takes account of a transit.  As we discuss in 
\S~\ref{sec:randomnoise},
$\Delta\chi_{\rm min}^{2}$
must be set sufficiently high to avoid spurious detections due to random noise.  

\begin{itemize}
\item To determine the dependence of $N_{p}$ on $L$, we note that in 
equation (\ref{equNt}), the only factor that depends on $L$ is $d_{max}^{3}$.  
For a particular star, the flux $f = L/(4 \pi d^{2})$, thus 
$d \propto L^{1/2}$, and so 
$N_{p} \propto L^{3/2}$.

\item For the dependence of $N_{p}$ on $R$, we note that 
$f \propto d^{-2}$, and that $\sigma \propto f^{-1/2}$, so 
$d \propto \sigma$.  From equation (\ref{equchi}), we see that 
$\sigma \propto \delta N_{t}^{1/2}$.  Since $N_{t}$ is the total number of 
observations of the star during transits over the length of the survey, 
$N_{t} = N_{\rm obs}(2 R)/(2 \pi a)$.  Also, $\delta = (\pi r)^{2}/(\pi 
R)^{2}$.  Thus, $d_{max} \propto R^{-3/2}$.  Combining these 
relations with the explicit factor of $R$ in equation (\ref{equNt}) 
itself, we arrive at $N_{p} \propto R^{-7/2}$.

\item For the dependence of $N_{p}$ on $a$, we see that $d \propto \sigma$, and 
$\sigma \propto \delta N_{t}^{1/2}$.  Using $N_{t} \propto a^{-1}$, 
we have $d_{max}^{3} \propto a^{-3/2}$, and so $N_{p} \propto a^{-5/2}$.

\item Finally, for the dependence of $N_{p}$ on $r$, the only factor that depends 
on $r$ is $\delta$, from $\delta = (\pi r)^{2}/(\pi R)^{2}$.  
Thus $\sigma \propto r^{2}$, and so $N_{p} \propto r^{6}$.
\end{itemize}

Consolidating the dependence of $N_{p}$ on the various parameters, we finally 
arrive at, 
\begin{equation} \label{equfin}
N_{p} = \frac{4}{3} \pi n \eta \, d_{0}^{3} \, \left( \frac{R_{0}}{a_{0}} 
\right) \left( \frac{L}{L_{0}} 
\right)^{3/2} \left( \frac{R}{R_{0}} \right)^{-7/2} \left( \frac{a}{a_{0}}
\right)^{-5/2} \left( \frac{r}{r_{0}} \right)^{6} ,
\end{equation}
where $d_0$ = $d_{max}(L_{0},R_{0},a_{0},r_{0},0)$.  That is, $d_{0}$ is the 
distance out to which a planet-star system with $i = 90^{\circ}$ and the fiducial 
parameters $L_{0}, R_{0}, a_{0}, r_{0}$ can just barely be detected at the S/N 
threshold.

We now seek to simplify equation (\ref{equfin}) by integrating over
the local stellar population at fixed absolute magnitude, $M_V$, and
so we replace the three independent variables $(n,R,L)$ by the single
variable $M_V$.  We consider two regimes $M_V \geq 6$ and
$M_V \leq 6$ (with one overlapping bin at $M_V = 6$, which we will
use later to check for consistency).

We first treat the fainter regime.  Here,
the main sequence is relatively narrow.  Hence, $R$ may be regarded
as a function of $L$ (and so, therefore, of $M_V$), while $n$ is
simply the number density of stars in a given magnitude bin.
Hence, the ``integration'' amounts to a simple multiplication of
factors.

We adapt the number density of stars $n$ from the empirically determined 
local stellar 
luminosity function (LF): for the range (9 $\leq$ $M_{V}$ $\leq$ 18) we use the  
LF reported in \citet{zheng01}, and for the range 
(6 $\leq$ $M_{V}$ $\leq$ 8) we use the LF of \citet{bessell93}.  

To estimate stellar radii, we combine the linear color-magnitude relation, 
$M_{V} = 3.37(V - I) + 2.89$ from \citet{reid91}, a color/surface-brightness 
relation $\log (R/R_{\odot}) = 0.69 \, + \, 0.2226(V-K) - 0.2M_{V}$, based on 
the data of \citet{belle99}, and $VIK$ color-color 
relations for dwarfs from \citet{bessell88}.  We can 
therefore calculate the relative number of systems with a fixed $a$ and $r$ 
as a function of $M_{V}$, which we designate
\begin{equation} \label{equF}
F(M_{V}) = \left[ \frac{n(M_{V})}{n_{0}} \right] \left[ 
\frac{L(M_{V})}{L_{0}} \right]^{3/2} \left[ 
\frac{R(M_{V})}{R_{0}} \right]^{-7/2} 
\end{equation}
where $n_{0}, L_{0},$ and $R_{0}$ are the normalizations chosen below.

For the upper main sequence, $M_V \leq 6$, we evaluate
$F(M_V)$ directly using the Hipparcos catalog \citep{hip97}.  For
example, the LF for $M_V=4$ would be computed by
summing $\sum_i [(4/3) \pi D_i^3]^{-1}$ over all stars within the
Hipparcos completeness limit, $V < 7.3$, having $3.5 <M_V <4.5$,
and lying within 50 pc.  The distance $D_i$ is the minimum of 50 pc
and the distance at which the star would have $V=7.3$.  Actually,
we are not directly interested in the LF at $M_V=4$, but rather in the
integral of $L^{3/2} R^{-7/2}$ over the subpopulations that make up
the $M_V = 4$ bin of the LF.  Hence
\begin{equation}
F(M_V=4) \, = \sum_{3.5<M_{V,i}<4.5}
\biggl({L_i\over L_0}\biggr)^{3/2}
\biggl({R_i\over R_0}\biggr)^{-7/2}
\biggl({4\over 3}\pi n_0 D_i^3\biggr)^{-1}.
\label{equLF}
\end{equation}
The stellar radii are determined from Hipparcos/Tycho $(B_T,V_T)$
photometry and the color/surface-brightness relation of \citet{gould03},
$\log {R/R_\odot} = 0.597 + 0.536(B_T - V_T) - {M_{V_T}/5}$,
ultimately derived from \citet{belle99}.

To normalize $F$, we adopt the values associated with $M_V = 5$ stars:
$L_0 = 0.86\,L_\odot$,
$R_0 = 0.97\,R_\odot$,
$n_0 = 0.0025\,{\rm pc}^{-3}$.

\begin{figure}[t]
\epsscale{1.0}
\plotone{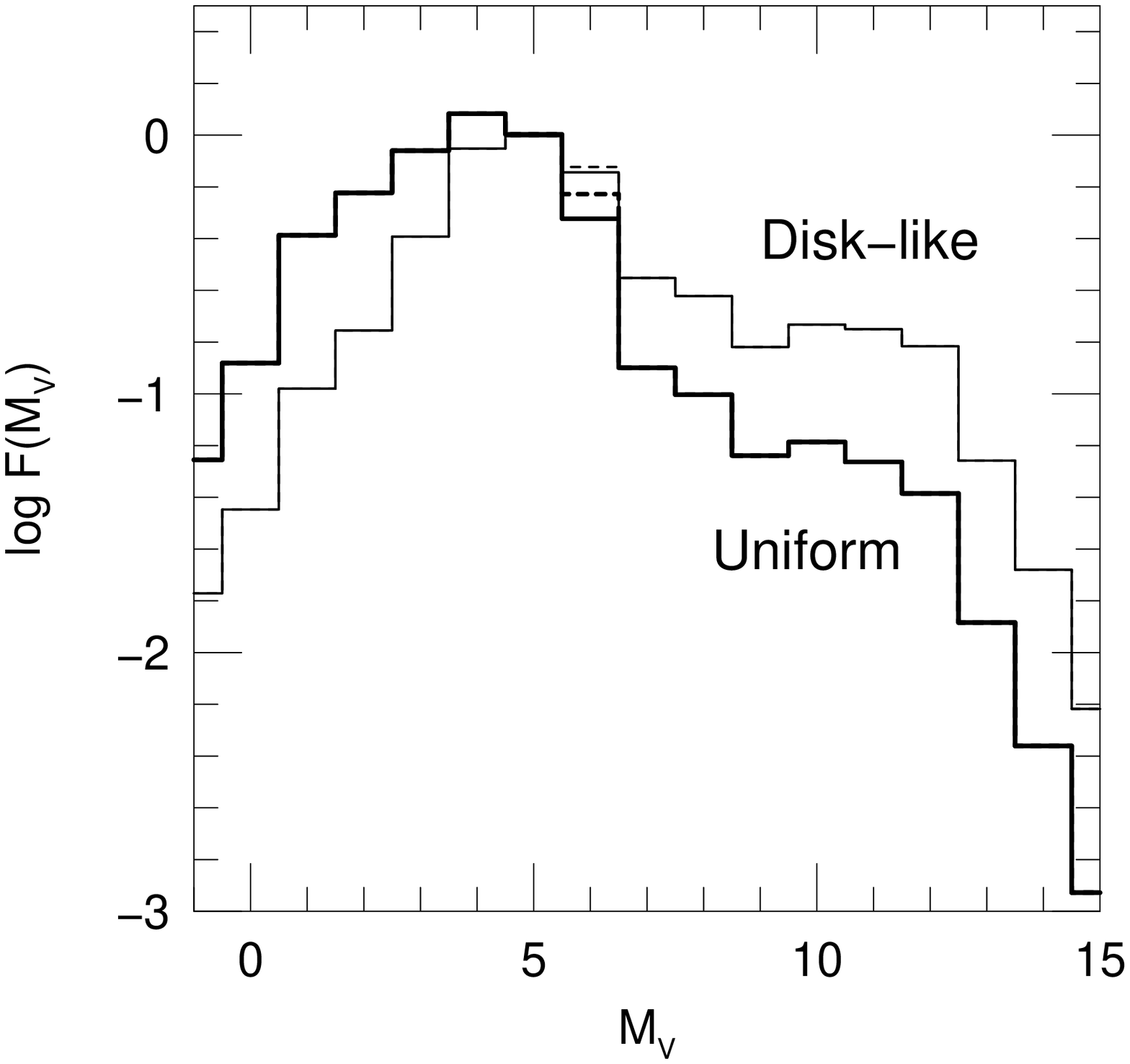}
\caption{The relative number of potential transiting systems $F(M_{V})$ probed for 
fixed planetary radius $r$ and semi-major axis $a$ as a function of $M_{V}$.  
The bold line applies to a uniform distribution of stars -- to model the 
immediate solar neighborhood.  The thin line applies to a thin disk -- to 
model a search of a large portion of the Galactic disk.  The dashed lines indicate 
where the two different methods for calculating the spatial density (as described 
in \S~\ref{sec:scalingrelations}) overlap in each case.  The distributions are 
arbitrarily scaled such that $F(M_{V}=5)=1$.}
\label{figF}
\end{figure}

The resulting function 
(Fig.~\ref{figF}) shows that the majority of stars that are probed will 
be F and G type ($2 \la M_{V} \la 6.5$).  To check how this distribution depends
on the size of the volume sampled, we recalculate the distribution function 
for the case in which the observed stars cover a much larger volume -- one which 
would be better described by a thin disk, rather than a spherically uniform 
distribution.  For this case we find that the scalings shown in equation 
(\ref{equF}) are replaced by $F(M_{V}) \propto n h L R^{-2}$, where $h$ is 
the scale height of each population. 
The distribution function is still dominated by F and G stars, but there are 
more K and early to mid M stars ($7 \la M_{V} \la 12$).  Of course, it is 
common knowledge that magnitude-limited $(F\propto n L^{3/2})$ samples of
main-sequence stars will be dominated by F and G stars.  The interesting feature
of Figure \ref{figF} is that this result does not qualitatively change despite
the addition of the factor $R^{-7/2}$ in equation (\ref{equF}), which very
strongly favors later-type stars.

\section{RANDOM NOISE} \label{sec:randomnoise}

Equation (\ref{equfin})
describes what kinds of XSP systems can be detected by a certain survey,
given a photometric detection limit.  The threshold $\Delta\chi^2_{\rm min}$ 
is determined by taking account of the fact that the data stream from an 
XSP search must be analyzed for any combination of the parameters 
$R$, $a$, and $r$ within reasonable ranges.  Such analysis will, 
however, yield a number of false-positive detections due to random noise.
The threshold value of $\Delta \chi_{\rm min}^{2}$ must be chosen to yield a
manageable number of candidate systems for follow-up observations.

To determine $\Delta \chi_{\rm min}^{2}$, we generate
1000 independent simulated streams of photometric observations of a single 
system with a host star of one solar mass and a circular planetary orbit.  We 
attempt to simulate a schedule that would be characteristic of an all-sky 
survey that re-images a given star approximately every few days with varying
intervals between observations.  For each data stream, we generate 1000 observations at 
irregular intervals over 1800 simulated days, and then phase
the observations for a range of periods from 3.0 to 3.1 days and a range of 
transit lengths.  The number of the observations and the duration of the 
simulated survey are arbitrarily chosen as plausible characteristics 
for the type of all-sky survey we envision.  

The duration of the transit depends on the orientation 
of the system with respect to our line of sight, and we analyze the data for 
eight equally spaced, progressively more inclined orientations.  For each 
orientation, we test for 16 equally spaced phases.  We test the resulting 
data sets for transit-like dips in the light curve, which we define simply 
as intervals during which the local mean light curve dips significantly below 
the global average.  This $(8\times 16)$ grid structure is chosen empirically:
we find that the number of false positives increases linearly with grid
density below this density and then flattens above it.

\begin{figure}[t]
\epsscale{1.0}
\plotone{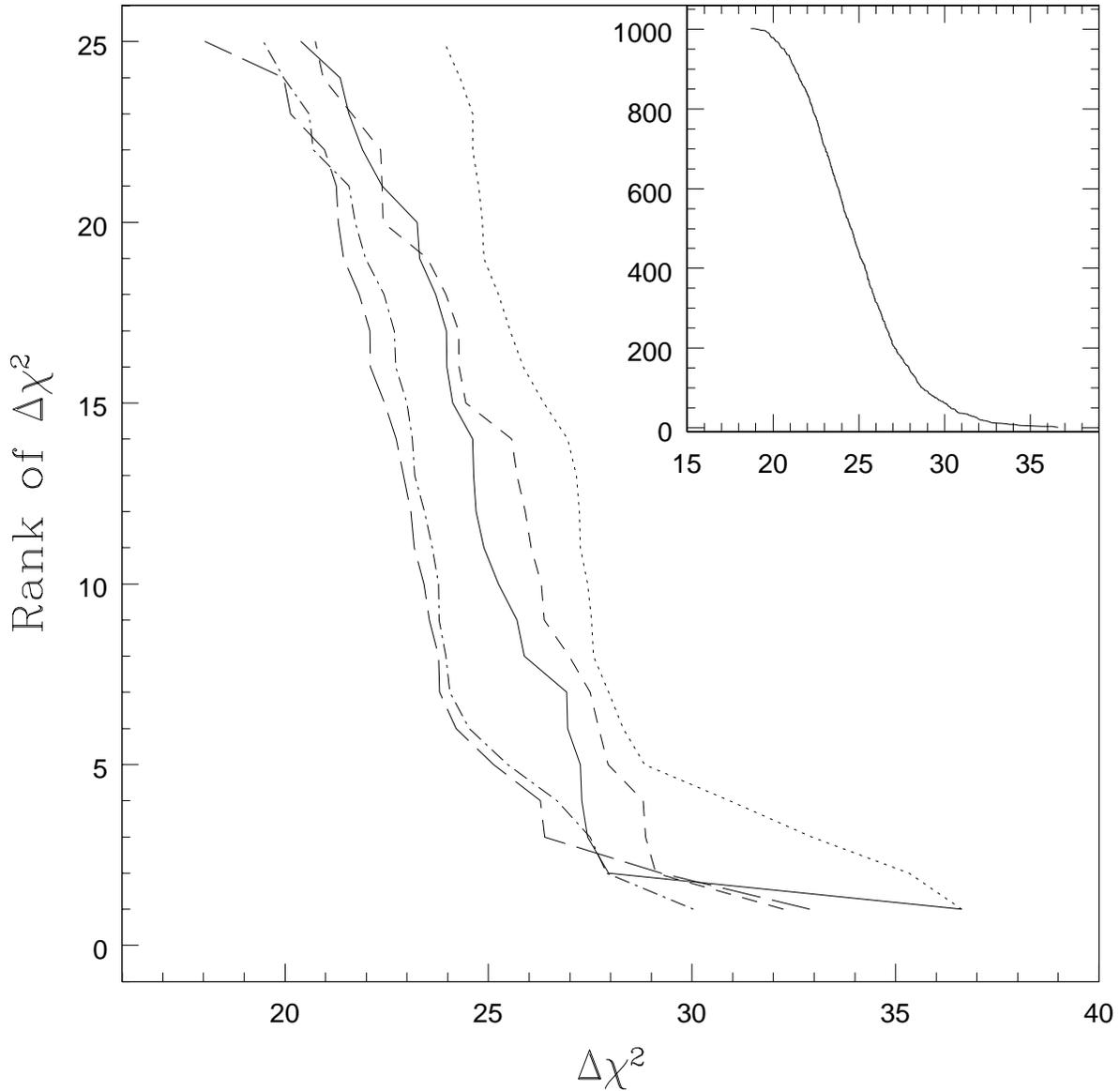}
\caption{The main plot shows the highest $\Delta \chi^{2}$ values for each of 25 runs, plotted by rank, for 
each type of observing schedule.  For four of the curves, a period search is 
conducted only for periods of 3.0 days to 3.1 days.  Shown are a random 
schedule (\it solid), \rm a regular schedule (\it short-dashed), \rm a regular 
schedule with 14.4-minute pair-spacing (\it long-dashed), \rm a regular schedule 
with 43.2-minute pair-spacing (\it dot-dashed).  \rm The dotted line is the random 
schedule with a period range of 3.0 to 3.5 days.  The inset plot shows the main simulation of 
1000 data sets.  The 0.1\% highest value is at 36.6.}
\label{figChi}
\end{figure}

The result shows that for this kind of system, 
in order to restrict follow-up analysis to the 
0.1\% of the full sample most likely to yield a true transit detection, 
the value for $\Delta \chi_{\rm min}^{2}$ should be set to $\sim$ 36.6.  We take 
the highest value of $\Delta \chi^{2}$ from each of the 1000 sets, and of those 
highest values, we then sort the 1000 sets from highest value of $\Delta \chi^{2}$ 
to lowest.  In 
Fig. \ref{figChi} (inset), we plot the highest value of $\Delta \chi^{2}$ 
for each of the 1000 data streams.

To check 
the robustness of this number, we run additional simulations of 25 data sets, each with different
observation schedules.  First, we generate a set of observations that are randomly 
distributed throughout a 1800 day mission, with the requirement that no two 
observations be less than 10 minutes apart.  Then we regenerate the data set 
with observations that are evenly spaced throughout the project.  We also 
conduct the analysis on data sets with evenly spaced pairs of observations, 
with the observations in each pair separated by 
14.4 minutes; and then again with the pair-spacing at 43.2 minutes.

These different configurations test for different types of observing schedules.  
For instance, a ground-based all-sky survey would most likely observe a star at 
essentially random times throughout a project.  On the other hand, a space-based 
mission with a slowly rotating telescope (similar to the Hipparcos satellite) 
would observe a star in pairs of observations separated by minutes.  For 
various reasons there could be a certain regularity imposed on either the 
space-based or ground-based observations.

We find that the value for $\Delta \chi_{\rm min}^{2}$ does not depend strongly 
on the observing schedule.  We determine this by taking the highest value of 
$\Delta \chi^{2}$ for each of the 25 independent data streams, 
sorting the highest values of $\Delta \chi^{2}$ from each set in the manner described above, 
and then plotting the results for each type of observing schedule 
(Fig. \ref{figChi}).  We find that there is only a $\sim$ 10\% variation in 
$\Delta \chi_{\rm min}^{2}$ for different schedules.
 
The value of $\Delta \chi_{\rm min}^{2}$ depends on the number of different 
parameter configurations that are tested.  When we rerun our analysis using a 
period range of 3.0 to 3.5 days, we increase the size of our parameter space 
by a factor of 5.  We expect that the value of $\Delta \chi_{\rm min}^{2}$ 
depends on the size of the parameter space, $\Psi$, according to 
$\Delta \chi_{\rm min}^{2}(\Psi_{new}) - 
\Delta \chi_{\rm min}^{2}(\Psi_{old}) = 
2 \ln (\Psi_{new}/\Psi_{old})$.  This prediction agrees with the simulations, 
which show that an increase in parameter space by a factor of 5 leads to an 
increase in $\Delta \chi_{\rm min}^{2}$ by $2 \ln (5) = 3.2$.

We then predict $\Delta \chi_{\rm min}^{2}$ if the size of the 
parameter space is expanded to search for transits with periods between 2 and 
10 days (the range of periods most probable for the sort of fiducial values of 
the other parameters we have chosen).  This would increase $\Psi$ by a factor 
of 80 compared to the period range of 3.0 to 3.1 days.  Therefore, 
$\Delta \chi_{\rm min}^{2}$ for the expanded range of periods is higher than 36.6 by 
$2 \ln (80) = 8.8$.  So the value expected for $\Delta \chi_{\rm min}^{2}$ 
for such an analysis is about 45.

\section{NUMBER OF SYSTEMS PROBED} \label{sec:sysprob}

The weak dependence of $\Delta \chi^2_{\rm min}$on the observing schedule
implies that the sensitivity of a project to planets essentially depends
only on the total number of photons detected from each star, and not on
the details of how they are collected.  For a given star, this number is 
obviously proportional to the flux.  We therefore characterize the sensitivity 
of the observing setup (telescope + detectors + duration + weather + etc.) by
$\gamma$, the total number of photons that are detected from a fiducial
$V = 10$ mag star during the entire project.  (Here we adopt $V = 10$  
as a reference, although the stars of interest lie in the range 
$8\la V\la 10$.)\ \  We can utilize the various relations used to 
derive equation (\ref{equfin}) to relate $\gamma$ to $V_{\rm max}$, the 
maximum apparent magnitude at which an equatorial transit can be detected,
\begin{equation} \label{equgam} 
V_{\rm max} = -2.5 \log \left( \frac{\Delta \chi_{\rm min}^{2} \pi a R^{3}}{r^{4} 
\gamma} \right) + 10 .
\end{equation}
Then, considering equations (\ref{equchi}), (\ref{equfin}), and (\ref{equF}), 
we obtain
\begin{equation} \label{equfin2}
N_{p} = 730 \, F(M_{V}) \left( \frac{a}{a_{0}} \right)^{-5/2} \left( \frac{r}{r_{0}} 
\right)^{6} \left( \frac{\gamma}{\gamma_{0}} \right)^{3/2} \left( \frac{\Delta 
\chi_{\rm min}^{2}}{45} \right)^{-3/2} ,
\end{equation}
where we have adopted $\gamma_{0} = 1.25 \times 10^{7}$, $a_{0}=10 \, R_{\odot}$, 
$r_{0} = 0.10 \, R_{\odot}$, and where we have made our evaluation at 
$M_{V} = 5$ (i.e. $R = 0.97 \, R_{\odot}$, $V_{\rm max} = 10$, $d_{0} = 100 \, $pc, 
and $n = 0.0025 \, {\rm pc}^{-3}$). Note that $\gamma_0= 1.25 \times 10^{7}$ 
corresponds to approximately 625 20-second exposures with a 5 cm telescope and a 
broadened $(V+R)$ type filter for one $V = 10$ mag fiducial star.

As mentioned above, a 
noteworthy feature of equation (\ref{equfin2}) is that $N_{p}$ depends on the 
characteristics of the survey primarily through the parameter $\gamma$.
Moreover, since $\Delta \chi_{\rm min}^{2}$ depends only logarithmically on the
size of the parameter space being explored, it plays a minimal role
in survey design compared to the other variables in equation (\ref{equfin2}).

We envision two scenarios to which these results will be applicable.  In one, 
a stream of photometric measurements from a space-based astrometric 
mission, such as GAIA, could be searched for transits.  
In the other, a ground-based survey could use one or more 
dedicated telescopes to search all bright stars in the solar neighborhood.  

\section{IMPLICATIONS FOR TELESCOPE DESIGN} \label{sec:ImpTelDes}

We now apply the general analysis of \S~\ref{sec:scalingrelations} and 
\S~\ref{sec:sysprob} to the problem of optimizing telescope design for
quickly locating a ``large'' number of bright ($V \la 10$) transiting systems.  Since
only one such systems is now known, we define ``large'' as ${\cal O}(10)$.
 From equation (\ref{equfin2}) and the 0.75\% frequency of hot jupiters
measured from RV surveys, there are roughly 5 such systems to be discovered
over the whole sky per magnitude bin at $M_{V} = 5$ for $V_{\rm max}=10$.  Hence, 
from Figure \ref{figF}, of order 25 are to be discovered from all spectral 
types.  It would, of course, be possible to discover even more by going 
fainter, but setting this relatively bright 
limit is advisable for three reasons.  First, as we argued in
the introduction, the brightest transits are the most interesting 
scientifically, and most of the transits detected in any survey will be
close to the magnitude limit.  Second, as we discuss below, a high dynamic range,
$\Delta V = V_{\rm max}-V_{\rm min}$, can only be
achieved at considerable cost to the observing efficiency.  Hence, if 
high efficiency is to be maintained, setting $V_{\rm max}$ fainter means 
eliminating the brightest (most interesting) systems.  Third, at 
$V_{\rm max}=10$, we are already reaching distances of 100 pc for G stars.
Hence the number of transits observed in fainter surveys will not continue
to grow as $d_0^3$ as in equation (\ref{equfin}).

In previous sections, we ignored the loss of sensitivity to systems
that are brighter than $V_{\rm min}$, which is set by saturation of the
detector (or more precisely, by the flux at which detector non-linearities
can no longer be accurately calibrated).  This fraction is $10^{-0.6\Delta V}$,
or 6\% for $\Delta V=2$, which we therefore adopt as a sensible goal.
That is, we wish to optimize the telescope design for,
\begin{equation}
8 = V_{\rm min} < V < V_{\rm max} = 10.
\label{equvminmax}
\end{equation}
(In any event, essentially all stars $V<8$ have already been surveyed
for XSPs using RV, and the problem of determining which among the 
planet-bearers have transits is trivial compared to the problem of conducting
an all-sky photometric variability survey.)

Optimization means maximizing the photon collection rate, $\gamma/T$, where
$T$ is the duration of the experiment and $\gamma$ is, again, the total
number of photons collected from a fiducial $V=10$ mag star.  Explicitly,
\begin{equation} \label{equGam1}
\gamma = K {\cal E} T D^2 \frac{(\Delta \theta)^2}{4\pi},
\end{equation}
where $\Delta\theta$ is the angular size of the detector,  $D$ is the 
diameter of the primary-optic, ${\cal E}$ is the fraction of the time
actually spent exposing, and $K$ is a constant that depends on the telescope,
filter, and detector throughput.  For our calculations, we assume
$K= K_0\equiv 40 \, e^-\,\rm cm^{-2}\,s^{-1}$, which is appropriate for a broad 
$(V+R)$ filter
and the fiducial $V=10$ mag star.  The design problems are brought into sharper
relief by noting that $\Delta\theta = {\cal L}/D{\cal F}$, where
${\cal L}$ is the linear size of the detector and ${\cal F}$ is the focal
ratio, or $f$/\#, of the optics.  Equation (\ref{equGam1}) then becomes
\begin{equation} 
\gamma = {K {\cal E} {\cal L}^2 T \over 4\pi{\cal F}^2}.
\label{equGam2}
\end{equation}

That is, almost regardless of other characteristics of the system, the 
camera should be made as fast as possible.  We will adopt ${\cal F}=1.8$, below
which it is substantially more difficult to fabricate optics.
A more remarkable feature of equation (\ref{equGam2}) is that all explicit 
dependence on the size of the primary optic has vanished: a 1 cm telescope
and an 8 m telescope would appear equally good!  Actually, as we now show,
there is a hidden dependence of ${\cal E}$ on $D$, which favors small 
telescopes.

\subsection{Considerations for Aperture Size}

The global efficiency ${\cal E}$ can be broken down into two factors,
${\cal E} = {\cal E}_0 {\cal E}_S$, where ${\cal E}_0$ is the fraction of
time available for observing (i.e., during which the sky is dark, the weather
is good, etc.), and ${\cal E}_S$ is the fraction of this available observing
time that the shutter is actually open.  The first factor is not affected
by telescope design and so will be ignored for the moment.  The second factor
should be maximized.  The smaller the telescope aperture is, the longer the 
exposures can be before a $V_{\rm min}=8$ mag star saturates.  Since the readout 
time is fixed, a smaller fraction of time is lost to read-out.
We adopt as benchmarks a detector with pixel size of $\Delta x_p=9\,\mu$m, well
depths of $10^{5} \,\, {\rm e}^{-}$, and a telescope diameter of 5 cm.

To make explicit calculations, $\theta_{\rm PSF}$, the
full width at half maximum of the point spread function (PSF) must be
specified.  For the fast optics (${\cal F} <$ few) we consider here, the diffraction 
limit is always much smaller than a pixel,
regardless of aperture:
$\theta_{\rm diff}/\theta_p \sim 1.22\,
{\cal F}\lambda/\Delta x_p \sim 0.16$ (for our fiducial choices).
At the small apertures we will consider, the diffraction limit is
larger than the seeing, so it is possible to make the PSF much
smaller than a pixel, $\theta_{\rm PSF}\ll \theta_p$.  This would
have the advantage of reducing sky noise and is a useful approach
when it is possible to always center the telescope at the same field
position as is the case for ``point and stare'' experiments.  However, for
an all-sky survey, which cycles through many fields, such precision
repeat pointing is extremely difficult.  Without it, precision
photometry is impossible unless the sub-pixel response of the CCD is
mapped out in detail.  We therefore adopt a Nyquist-sampled PSF, for
which the sky noise is approximately that falling on $4\pi\sim 13$
pixels.

Our overall consideration for telescope design must take into account three 
factors.  First, we with to maximize observing efficiency ${\cal E}_S$.  Second, 
we wish to achieve the highest possible signal-to-noise ratio.  Third, we must 
avoid any distortion problems with the optics.  There are four effects through 
which aperture size can impact these factors.  Two of these effects, observing 
efficiency and scintillation noise, will drive us to larger telescopes, while 
the other two effects, sky noise and focal plane distortion, will drive us to 
smaller telescopes.  As we show below, for the observing parameters we have 
specified an aperture of 5 cm ensures a manageable (and unique) balance 
between the various effects.

\subsubsection{Exposure Time vs. Readout Time}   \label{subsubsec:exptime}

Assuming Nyquist sampling, at most half the light from a point source 
falls within one pixel.  We can directly calculate the ratio of time lost to readout 
$T_{\rm read}$ to the time spent exposing $T_{\rm exp}$,
\begin{equation}
\frac{T_{\rm read}}{T_{\rm exp}} = 1 \left( \frac{D}{5 \, \rm cm} \right)^{2} 
10^{-0.4(V_{\rm max}-10)}
\left( \frac{W}{W_{0}} \right)^{-1} \frac{K}{K_{0}} \frac{T_{\rm read}}{30 {\rm s}} 
10^{0.4(\Delta V - 2)},
\label{equtexp}
\end{equation}
where $W$ is the well depth of the detector pixels, and $W_{0}$ = $10^{5}{\rm e}^{-}$ 
is a fiducial well depth.  
Note that the factor $10^{-0.4(V_{\rm min}-10)}$, which arises from the need to 
avoid saturation of the brightest stars (where $V_{\rm min} = V_{\rm max} - \Delta V$), 
has been broken up into two terms to permit easy comparisons of equation 
(\ref{equtexp}) with equations (\ref{equscint2}) and (\ref{equskynoise}) below.

In order to maximize the efficiency ${\cal E}_{S}$, the fraction of observing 
time devoted to readout should be minimized, and therefore, 
according to equation (\ref{equtexp}), so should the aperture size.  
The telescope will 
operate reasonably efficiently so long as $T_{\rm read} \la T_{\rm exp}$.

\subsubsection{Scintillation}   \label{subsubsec:scint}

Another concern that arises for small apertures is the effect of atmospheric 
scintillation, which is characterized by (\citealt{young67}; \citealt{war88}),
\begin{equation}
\frac{\Delta I}{I} = S_{0} \left( \frac{D}{\rm cm} \right)^{-2/3} X^{3/2} 
\exp \left( - \frac{h}{h_{0}} 
\right) \left( \frac{2 T_{\rm exp}}{\rm sec} \right)^{-1/2} ,
\label{equscint1}
\end{equation}
where $S_{0}$ = 0.09, $X$ is the 
airmass, $h$ is the altitude of the observatory, and $h_{0} = 8$ km is the 
scale height of the atmosphere.  Using values of 
$D = 5$ cm, $X = 1.5$, $h = 2$ km, and $T_{\rm exp} = 30$ s, we find 
$\Delta I / I = 0.0057$.

Hence, for exposure times set by a saturation threshold, 
$T_{\rm exp} \propto D^{-2}$ (see eq. [\ref{equtexp}]), 
\begin{equation}
\frac{\rm Scintillation \, \, Noise}{\rm Source \, \, Noise} = 
1 \left( \frac{D}{5 \, \rm cm} \right)^{1/3} 10^{-0.2(V - 10)} .
\label{equscint2}
\end{equation}
Therefore, despite the common perception that 
scintillation is a greater problem for smaller telescopes, for the fixed 
photon counts per exposure that are of interest in the present context, 
scintillation noise increases with {\it increasing} aperture.  However, this
dependence is fairly weak.  For $D = 5$ cm, the scintillation noise is just 
slightly smaller than the photon noise at $V = 10$. 

\subsubsection{Sky Noise}   \label{subsubsec:skynoise}

All the calculations in \S\S~\ref{sec:scalingrelations}, \ref{sec:randomnoise}, 
and \ref{sec:sysprob} have assumed that sky noise is 
negligible, i.e. $V \ll V_{\rm sky}$ where $V_{\rm sky}$ is the light from the 
sky falling on 13 pixels (for Nyquist sampling).  Assuming a somewhat 
conservative mean sky brightness of $V=20.0\,\rm mag/arcsec^{2}$, one finds
\begin{equation}
V_{\rm sky} = 10.6 + 5\log \left[
{D\over 5\,\rm cm}\,{{\cal F}\over 1.8} \, \left( \frac{\Delta x_{p}}{9\,\rm \mu m}
\right)^{-1}
\right].
\label{equsky}
\end{equation}
To determine how much of a problem sky noise will be, we can compare it to the 
amount of photon noise,
\begin{equation}
\frac{\rm Sky \, \, Noise}{\rm Source \, \, Noise} = \frac{3}{4} 
\left( \frac{D}{5 \, \rm cm} \right)^{-1} 10^{0.2(V-10)} .
\label{equskynoise}
\end{equation}
Thus, sky noise is less than photon noise for a 5 cm telescope, but would become 
a serious problem for a substantially smaller aperture.

\subsubsection{Focal Plane Distortion}  \label{subsubsec:focalplane}

Focal plane distortions toward the edge of the detector become difficult 
(and expensive) to correct when the field of view is too large.  For example, 
for a $4k\times 4k$ detector with $\Delta x_p=9\,\mu$m pixels, 
\begin{equation}
\Delta \theta = 23^\circ \left( \frac{D}{5\,{\rm cm}} \right)^{-1} 
\left( \frac{\cal F}{1.8} \right)^{-1} \, .
\label{equ:focalplane}
\end{equation}
Again, this is manageable for $D = 5$ cm but could would potentially be a 
problem for smaller apertures.

Although this field of view is not so large as to create focal plane distortions, 
it is important to note that for a field of view this large, the telescope must be 
placed on an equatorial mount.  For the alternative, an alt-az mount, the rotation 
of the sky will cause stars at one edge of the 
field to move faster across the detector than stars at the opposite edge.  For
similar reasons, the telescope must track rather than using drift scan.

\subsection{Optimal Telescope Design} \label{sec:OptTelDes}

The relationships described in \S~\ref{subsubsec:exptime} through 
\S~\ref{subsubsec:focalplane} can be used to optimize the aperture $D$ 
for any survey parameterized by a given $V_{\rm max}$.  One is driven to 
smaller apertures by the goals of
minimizing scintillation noise and the fraction of time spent on readout, and 
to larger apertures by the goals of minimizing sky noise and field distortion.
Given available ${\cal L}=3.6\,$cm $4096\times 4096$ detectors and
reasonably fast ${\cal F}=1.8$ optics, a $D=5\,$cm telescope is
optimal for an all-sky survey of $V=10$ stars.  Among all existing transit 
programs of which we are aware, the WASP
telescope ($D=6$ cm lens, ${\cal F}=2.8$ focal ratio,
$2k \times 2k$, ${\cal L}=3\,$cm detector, \citealt{str02}) 
comes closest to meeting these design specifications.

We use equation (\ref{equGam2}) to determine the required duration of the 
experiment using our optimally-designed telescope, adopting $\gamma=\gamma_0$, 
$K_0 = 40 \, {\rm e}^-\,\rm cm^{-2}\,s^{-1}$, 
${\cal E}_{0}=20\%$, ${\cal E}_{S}=50\%$, ${\cal L}=3.6\,$cm, and ${\cal F}=1.8$.
The total time required to conduct 
the survey assuming only source photon noise is $T=4\,\rm months$.  Taking 
into account sky noise and scintillation increases that time by a factor 
of 2.5 to 10 months.

For surveys that intend to search for transits of stars fainter 
than the range we consider in this paper ($8 = V_{\rm min} < V < V_{\rm max} = 10$), 
the relations in \S~\ref{subsubsec:exptime} through 
\S~\ref{subsubsec:focalplane} operate somewhat differently.  For the 
magnitude range we are considering, equations (\ref{equtexp}) and 
(\ref{equscint2}) provide an upper limit on the aperture size, while 
equations (\ref{equskynoise}) and  (\ref{equ:focalplane}) place a 
lower limit, in which both limits converge to $D = 5 {\rm cm}$.  At 
fainter magnitudes, the aperture size is restricted according to 
equations (\ref{equtexp}) and (\ref{equskynoise}):
\begin{equation}
10^{0.2(V - 10)} \la \frac{D}{5 \, {\rm cm}} \la 10^{0.2(V_{\rm max} - 10)}.
\label{equ:highmag}
\end{equation}
This equation shows how the aperture size limits combine into a single 
scaling relation.  What of the other two limiting factors?  The issue of focal
 plane distortions is unimportant at fainter magnitudes, since 
$\Delta \theta < 23^\circ$, which is outside of the regime where such 
distortions are a factor.  Scintillation effects are also not a factor 
at fainter magnitudes, since the scintillation-noise restriction requires 
that $(D/5 \, {\rm cm}) < 10^{0.6(V-10)}$, which is a looser restriction than 
that of equation (\ref{equ:highmag}).

Therefore, for a survey of transits at a magnitude range 
$8 = V_{\rm min} < V < V_{\rm max} = 10$, the aperture size limits converge 
to $D = 5 \, {\rm cm}$, while at fainter magnitudes, the aperture size is 
given by equation (\ref{equ:highmag}).

\subsection{Practical Implementation}

In the previous section, we estimated the duration of observations 
required to achieve the minimum S/N to detect transits by hot
Jupiters assuming certain fiducial parameters of a ground-based
telescope, but calculated within the framework of a literal ``all-sky''
($4\pi$) survey of randomly-timed observations that is more characteristic
of satellite missions.  The resulting estimate is useful for judging
the viability of a given observing setup, but it glosses over a 
key issue in the detection of transits, namely the problem of folding
the data.  As discussed in \S \ref{sec:randomnoise}, the number of folds (and hence
the size of the search space that must be probed) scales directly
as the duration of the experiment.
This larger search space increases both the minimum $\Delta\chi^2$
for a robust detection and the amount of computing power needed to
sift through the search space.  The first effect is logarithmic in
the size of the search space, so a 10-fold increase changes $\Delta \chi_{\rm min}^2$
by only $\sim \ln 10\sim 2.3$, which is well under 10\%.  However,
the second effect is linear in the search space and so could easily
overwhelm available computing resources if not carefully monitored.
That is, there are important drivers for keeping the duration of
observations of any given field to a minimum and hence for exploring
the question of whether it is better to break up the ``all-sky''
survey into several smaller components, each of which could be
completed in a shorter time.  Indeed, the OGLE experiment \citep{udal02},
the only transit experiment to successfully detect a transit \citep{kon03}, was
motivated by these considerations to concentrate its observations
over a month duration so as to limit the number of foldings.

To estimate the true duration of the project required the achieve the 
minimum S/N, one must take account of two factors: First, as a 
practical matter, ground-based surveys from a single location can cover
only an angular area $\Omega<4\pi$.  Second, during a year of continuous
observations, any given patch of sky is observed only for about 6 months.  
These factors change the previous estimates from \S \ref{sec:OptTelDes}.
The first factor implies that the observing time required to reach
minimum S/N is actually lower than the time calculated from equation 
(\ref{equGam2}) by a factor of $\Omega/4\pi\sim 0.5$, since the project 
is observing about half the angular area on the sky.  The impact of the 
second factor is more complex.  In a given night, only half the 
accessible angular area $\Omega$ can be observed.  Therefore, each night 
the available area to observe is lower by an additional factor of 2, which 
means the rate of observations of a given point on the sky will double 
again.  There are two scenarios at this point.  The first scenario is 
that after accounting for the first factor, the time required for 
sufficient S/N is less than a year.  In this case, the two factors 
combine to lower the estimate from \S \ref{sec:OptTelDes} by a factor of 
about 4.  In the second scenario, the time required for sufficient S/N 
is more than a year.  In this case, after observing for 6 months the 
minimum S/N is not reached, and therefore the project will have 
to pick up again six more months later when the target is again visible, 
and so the second factor does not apply.  As we see below, however, for 
the parameters we are considering, we are well within the regime of the 
first scenario.

Applying the first factor to the result in \S \ref{sec:OptTelDes} reduces the calculated
time from 10 months to 5 months.  Since this latter duration
is indeed smaller than 1
year, the second factor implies that the requisite S/N will actually be
reached in 2.5 months.  This is only slightly longer than the duration
of the OGLE observations.

However, real experiments inevitably have larger errors than expected.
(We mention one source of additional errors below.)  If the errors
prove sufficiently larger that the experiment requires more than 1 year (after application
of the first factor) to achieve the minimum S/N, then one would
not gain the advantage of the second factor.  In this case, it would
be better to break the sky up into strips by declination, and observe
each strip for a year, so as to increase the S/N obtained in each strip
during a single year, and so to permit the application of the second
factor.

Another real-world consideration is that the photometric errors
will not be Gaussian.  For example, the 
Sloan Digital Sky Survey (SDSS) photometry errors, while Gaussian
in their core, deteriorate to an exponential profile in the wings
beginning at about $3\,\sigma$ \citep{ive03}.  Such deterioration is
likely to set in earlier for the small-aperture, wide-field, low-budget
cameras that we envisage here.  The non-Gaussian {\it form} of the
errors has no practical impact on our analysis: there are so many
data points that the central limit theorem guarantees
that their combined behavior in each phase bin will be Gaussian
(as we have implicitly assumed in \S \ref{sec:randomnoise}.  However, the non-Gaussian
tails will tend to increase the $\sigma$ of the distribution relative
to what would be inferred from the core, which of course will degrade
the sensitivity of the experiment.  If the errors are as well-behaved
as those of SDSS, this problem can easily be resolved by the standard
device of 3-$\sigma$ clipping.  If not, then more complex strategies
will be required.  These are likely to be among the biggest practical
problems facing the analysis, but in the absence of real data, they 
cannot be further analyzed here.

\acknowledgments This work was supported in part by
grant AST 02-01266 from the NSF.

\begin{thebibliography}{}

\bibitem[Bessell \& Brett(1988)]{bessell88}  

\{Bessell, M. S. and Brett, J. M.\}\{1988\}\{Publ. Astron. Soc. Pac.\}\{100\}\{1134\}

\bibitem[Bessell \& Stringfellow(1993)]{bessell93}  

\{Bessell, M. S. and Stringfellow, G. S.\}\{1993\}\{Annu. Rev. Astron. Astrophys.\}\{31\}\{433\}

\bibitem[Brown \& Charbonneau(1999)]{brown99}  

\{Brown, T. M. and Charbonneau, D.\}\{1999\}\{Bull. Am. Astron. Soc.\}\{31\}\{1534\}

\bibitem[Burke et al.(2002)]{burke02}  

\{Burke, C. J., DePoy, D. L., Gaudi, B. S., Marshall, J. L. and Pogge, R. W.\}\{2002\}\{Astron. Soc. Pac. Conf. Ser.(in print)\}\{astro-ph/0208305\}

\bibitem[Charbonneau et al.(2000)]{char00}  

\{Charbonneau, D., Brown, T. M., Latham, D. W. and Mayor, M.\}\{2000\}\{Astrophys. J.\}\{529\}\{L45\}

\bibitem[Charbonneau et al.(2002)]{char02}  

\{Charbonneau, D., Brown, T. M., Noyes, R. W. and Gilliland R. L.\}\{2002\}\{Astrophys. J.\}\{568\}\{377\}

\bibitem[Cody \& Sasselov(2002)]{cody02}  

\{Cody, A. M. and Sasselov, D. D.\}\{2002\}\{Astrophys. J.\}\{569\}\{451\}

\bibitem[ESA(1997)]{hip97} 

\{European Space Agency\}\{1997\}\{The Hipparcos and Tycho Catalogues\}

\bibitem[Gould \& Morgan(2003)]{gould03}  

\{Gould, A. and Morgan, C. W.\}\{2003\}\{Astrophys. J.\}\{585\}\{1056\}

\bibitem[Henry et al.(1999)]{hen99}  

\{Henry, G., Marcy, G. W., Butler, R. P., Vogt, S. S. and Apps, K.\}\{1999\}\{IAU Circ.\}\{7307\}

\bibitem[Howell et al.(2000)]{how00}  

\{Howell, S. B., Everett, M., Davis, D. R., Weidenschilling, S. J., McGruder, C. H., III, and Gelderman, R.\}\{2000\}\{Bull. Am. Astron. Soc.\}\{32\}\{3203\}

\bibitem[Hui \& Seager(2002)]{hui02}  

\{Hui, L. and Seager, S.\}\{2002\}\{Astrophys. J.\}\{572\}\{540\}

\bibitem[Ivezic, Z. et al.(2003)]{ive03}%

\{Ivezic, Z., et al.\}\{2003\}\{Mem. Soc. Astron. Ital.\}\{20\}\{1\}

\bibitem[Konacki et al.(2003)]{kon03}%

\{Konacki, M., Torres, G., Jha, S., and Sasselov, D. D.\}\{2003\}\{Nature\}\{421\}\{507\}

\bibitem[Mallen-Ornelas et al.(2001)]{mal01}  

\{Mallen-Ornelas, G., Seager, S., Yee, H. K. C., Minniti, D., Gladders, M. D., Ellison, S., Brown, T. and Mallen, G. M.\}\{2001\}\{Bull. Am. Astron. Soc.\}\{199\}\{6602\}

\bibitem[Pojmanski(2000)]{poj00}  

\{Pojmanski, G.\}\{2000\}\{Acta Astron.\}\{50\}\{177\}

\bibitem[Reid(1991)]{reid91} 

\{Reid, N.\}\{1991\}\{Astron. J.\}\{102\}\{1428\}

\bibitem[Street et al.(2000)]{str00}  

\{Street, R. A. et al.\}\{2000\}\{Astron. Soc. Pac. Conf. Ser.\}\{219\}\{572\}

\bibitem[Street et al.(2002)]{str02}  

\{Street, R. A. et al.\}\{2002\}\{Astron. Soc. Pac. Conf. Ser. (in press)\}\{astro-ph/0209188\}

\bibitem[Udalski et al.(2002)]{udal02}  

\{Udalski, A.\}\{2002\}\{Acta Astron.\}\{52\}\{115\}

\bibitem[van Belle(1999)]{belle99}  

\{van Belle, G. T.\}\{1999\}\{Publ. Astron. Soc. Pac.\}\{111\}\{1515\}

\bibitem[Warner(1988)]{war88}  

\{Warner, B.\}\{1988\}\{High Speed Astronomical Photometry\}\{Cambridge University Press\}

\bibitem[Young(1967)]{young67}  

\{Young, A. T.\}\{1967\}\{Astron. J.\}\{72\}\{747\}

\bibitem[Zheng et al.(2001)]{zheng01}  

\{Zheng, Z., Flynn, C., Gould, A., Bahcall, J. N. and Salim, S.\}\{2001\}\{Astrophys. J.\}\{555\}\{393\}

\end{thebibliography}
\end{document}